\documentclass[letterpaper, 10 pt, conference]{ieeeconf}

\usepackage{amsmath}
\usepackage{amsfonts}
\usepackage{theorem}
\usepackage{graphicx}
\usepackage{epstopdf}
\usepackage{color}
\usepackage{float}
\usepackage{subcaption}

\usepackage{pgfplots}
\pgfplotsset{compat=newest}
\usetikzlibrary{plotmarks}
\usetikzlibrary{arrows.meta}
\usepgfplotslibrary{patchplots}
\usepackage{grffile}

\newtheorem{theorem}{Theorem}
\newtheorem{lemma}{Lemma}

\newcommand{\R}{\mathbb R}

\newcommand{\U}{\mathcal U}

\newcommand{\A}{\mathcal A}
\newcommand{\X}{\mathcal X}

\title{Geometric Motion Planning for Affine Control Systems with Indefinite Boundary Conditions and Free Terminal Time}
\author{Shenyu Liu, \and Yinai Fan  \and Mohamed-Ali Belabbas}

\begin{document}

\maketitle
\begin{abstract}
    The problem of motion planning for affine control systems consists of designing control inputs that drive a system from a well-defined initial to final states in a desired amount of time. For control systems with drift, however, understanding which final states are reachable in a given time, or reciprocally the amount of time needed to reach a final state, is often the most difficult part of the problem. We address this issue in this paper and introduce a new method to solve motion planning problems for affine control systems, where the motion desired can have indefinite boundary conditions and the time required to perform the motion is free. The method extends on our earlier work on motion planning for systems without drift. A canonical example of parallel parking of a unicycle with constant linear velocity is provided in this paper to demonstrate our algorithm.

%   {\color{blue} this is unclear at this stage:} We first show that the deficiency of boundary conditions can be completed by constraints on the Lagrangian with respect to the the derivative of the corresponding states instead, which can be verified via a perturbation argument.
    
%     On the other hand when the terminal time is free, a time scaling can be applied, which results in an augmented affine system with drift and indefinite boundary conditions. Thus use the techniques we developed in the previous work and the analysis on indefinite boundary conditions, we are able to plan the motion when terminal time is free. In the end of the paper we also provide a canonical example on unicycle with constant linear velocity to demonstrate our algorithm.
\end{abstract}

\section{Introduction}
%{\color{blue} The most important part here would be to clearly illustrate the issue for systems with drift: perhaps move up the example of the gymnast given at the end?}

Due to its ubiquity in control applications ranging from robotics to autonomous wheeled vehicles, motion planning has been widely studied (see, e.g., ~\cite{Laumond1998,lavalle_2006} and the reference therein) and a host of methods have been developed. 
%One of the control papers in which the issue of motion planning for non-holonomic systems was clearly delineated is the seminal paper~\cite{Brockett1982}, where finding a subRiemannian geodesic can easily be seen as solving a non-holonomic motion planning problem. For a more recent survey of this line of work, we refer to the recent monograph~\cite{Jean2014}. A common approach to non-holonomic motion planning is to use sinusoidal control function to, roughly speaking, generate the ``Lie bracket'' directions. See for example~\cite{li2012nonholonomic}.
From a theoretical point of view, the Chow-Rashevski Theorem provides us with conditions under which a driftless non-holonomic system is controllable~\cite{Chow1940}. the idea of \emph{small time local controllability} can be generalized to affine systems with drift as studied in \cite{Sussmann_STLC}, \cite{261504}. Beyond that the motion planning is still quite challenging and even the general case of controllability of nonlinear systems is still an open problem. Nevertheless, for some specific nonlinear systems with drift, motion planning or control algorithms are given in \cite{Luca1995,611856,788533,Zeng2019IterativeOC}.

Inspired by the curving shortening flow as studied in \cite{curvebook2001}, we have proposed a geometrical approach for motion planning in \cite{7963599} for driftless affine systems. The method works by ``deforming'' an arbitrary path between a given initial state $x^i$ and a given final state $x^f$ into an almost admissible trajectory for the system, from which we can extract the controls $u^*$ that drive the system from $x^i$ to $x^f$ approximately. Using a variational approach, we provided a proof of convergence of the method as well. By encoding obstacles into barrier functions, we are able to plan the motion while avoid obstacles. In the later work \cite{SL19NOLCOS}, we modified the algorithm so it is capable of addressing motion planning problems for affine systems with drift and input constraints by solving the so called \emph{affine geometric heat flow} equation.

Nevertheless, most literature mentioned above including our previous works only focused on motion planning problems with both end points of the motion as well as the time span fixed. In practice, we very often allow \emph{indefinite boundary conditions} (IBCs) and \emph{free terminal time} (FTT). It is appreciated that in many motion planning problems there is no need to fix the boundary conditions as a priority, or the optimal/feasible boundary states need to be determined in the motion planning problems. For example, consider the task of somersault performed by a robot made of linkages as studied in \cite{4209291}, \cite{7759702}. While the final joint angles are prescribed, there are no constraints on the final velocities of the joints if ``landing with impact" is allowed. On the other hand, if the final dynamic configuration is fully fixed for the same somersault problem, we have implicit constraints on choosing the initial configuration because of the conservation of angular momentum in mid-air and hence it is better to ask the algorithm to find the feasible initial configuration \cite{YF19WROCOS}. In terms of terminal time, it is less important for driftless systems because the time span can always be adjusted by input scaling; however, scaling will not work if there is drift or the control is constrained. In other words, the reachable space depends on the terminal time, which affects the feasibility of our motion planning problem so FTT is indeed crucial in our case and the terminal time cannot be fixed prior to the motion planning problem either. 

By deploying some ideas from calculus of variation and state augmentation, in this work we modify our previous algorithm so that it is able to solve motion planning problems with both IBCs and FTT. The paper has 6 sections. In Section~\ref{sec:prelim} our motion planning problem is formulated with necessary preliminaries. We then state the changes in our algorithm for handling IBCs in Section~\ref{sec:IBCs}, followed by the discussion of FTT in Section~\ref{sec:FTT}. Our algorithm is then examined in a case study on unicycle in Section~\ref{sec:example} and the paper is concluded in Section~\ref{sec:conclusion}.

\section{Problem formulation and preliminaries}\label{sec:prelim}
%For a vector $v\in \R^n$, define $v_{i:j}:=\begin{pmatrix}v_i&v_{i+1}&\cdots&v_j\end{pmatrix}$; that is, the extraction of the $i$-th to the $j$-th elements of $v$.

Consider an affine control system with drift:
\begin{equation}\label{sys}
    \dot x=h(x)+F(x)u
\end{equation}
Where $x\in\R^n$ are the states, and for each fixed $t$ and $x$, $h(x)\in\R^{n}$ is the drift and $F(x)\in\R^{n\times m}$ consists of columns of control directions and $u(t)\in\R^{m}$ is the control input. We further assume $m<n$ so the system is under-actuated and $F(x)$ is full rank for all $x\in\R^n$. Let $T>0$ be the terminal time, either free or fixed. For simplicity we assume that both the functions $h(\cdot),F(\cdot)$ are smooth and the input function $u(\cdot)\in L^2([0,T]\to\R^m)=:\U$. For a curve $x(\cdot)\in C^1([0,T]\to\R^n)=:\X$, Let $L$ be some Lagrangian defined with respect to $x(t),\dot x(t)$. Defined the \emph{action functional} $\A$ over $\X$ as follows:
\begin{equation}\label{def:A}
    \A(x):=\int_0^TL(x(t),\dot x(t))dt
\end{equation}
It is well known that Euler-Lagrange equation
\begin{equation}\label{eqn:EL}
    \frac{\partial L}{\partial x}(x^*(t),\dot x^*(t))-\frac{d}{dt}\frac{\partial L}{\partial \dot x}(x^*(t),\dot x^*(t))=0\quad \forall t\in[0,T]
\end{equation}
 is a necessary condition for a curve $x^*$ to be a local optimizer of \eqref{def:A} in $\X$. Instead of solving the system of ODEs \eqref{eqn:EL} directly, we can view the solution of \eqref{eqn:EL} as the steady state solution of a system of PDEs by
adding one more variable $s$ to the curve $x$ so now it becomes $x(t,s)$, defined on $[0,T]\times[0,\infty)$. Notice that for each fixed $s$, $x(\cdot,s)$ still represents a curve in $\X$. As $x(t,s)$ is a multivariable function, we use $x_t$ and $x_s$ to represent $\frac{\partial x}{\partial t},\frac{\partial x}{\partial s}$, respectively. We define the \emph{affine geometric heat flow} (AGHF) equation as follows:
\begin{equation}\label{eqn:HFE}
    x_s=G(x)^{-1}\left(\frac{d}{dt}\frac{\partial L}{\partial x_t}(x,x_t)-\frac{\partial L}{\partial x}(x,x_t)\right),
\end{equation}
where $G$ is a positive definite $n\times n$ matrix which will be given later. Unlike the case we have studied in our earlier work that both ends of the curve is fixed in the sense that there exists $x^i,x^f\in\R^d$ such that we have the boundary conditions 
\[
    x(0,s)=x^i,x(T,s)=x^f\quad\forall s\geq 0,
\]
we now only fix part of them in the case of IBCs. To be more precise, let $S_{bc}\subset S:=\{1,\cdots,d\}\times\{0,T\}$.  For some given $x^{bc}(i,t)\in\R^n,(i,t)\in S_{bc}$, 
\begin{equation}\label{partial bc1}
    x_i(t,s)=x^{bc}(i,t)\quad\forall (i,t)\in S_{bc}, s\geq 0
\end{equation}
We allow the case that $S_{bc}=\emptyset$, meaning that there are no boundary conditions at all. Meanwhile, in order to solve the parabolic type PDE \eqref{eqn:HFE}, we still need to feed it with an initial curve
\begin{equation}\label{ic}
    x(t,0)=z(t),\quad t\in[0,T]
\end{equation}
where $z(\cdot)\in\X'$  with
\[
\X':=\{x\in \X:x_i(t)=x^{bc}(i,t) \quad\forall (i,t)\in S_{bc}\}
\]
In other words, $\X'$ is the set of all $C^1$ curves with the IBCs satisfied. Note that the curves in $\X'$ do not need to meet the system dynamics \eqref{sys}. We also define the set of \emph{admissible paths} by
\[
\X^*:=\{x\in\X':\mbox{\eqref{sys} is satisfied for some }u\in \U\}    
\]
We say the motion planning problem is \emph{feasible} when $\X^*\neq \emptyset$. While finding a curve in $\X'$ is trivial, finding an admissible path in $\X^*$ is difficult. In this work we will show an efficient algorithm that gives an approximation to some $x^*\in \X^*$ when the problem is feasible.

\section{Indefinite boundary conditions}\label{sec:IBCs}

We first consider the case when the terminal time $T$ is fixed. Notice that when $S_{bc}=S$, we recover the fully constrained boundary conditions. However, when $S_{bc}$ is a proper subset of $S$, algebraically there are not enough boundary conditions in \eqref{partial bc1} so the ODEs \eqref{eqn:EL} are under-determined. Nevertheless, For being a minimizer of the functional \eqref{def:A}, it should also give $0$ variation with respect to perturbations in the free boundary states; In other words, we should also have
 \begin{equation}\label{partial bc2}
     \frac{\partial L}{\partial (x_t)_i}(x(t,s),x_t(t,s))=0\quad\forall (i,t)\in S\backslash S_{bc},s\geq 0
 \end{equation}
 \eqref{partial bc2} together with \eqref{partial bc1} are the new boundary conditions we will use for solving the AGHF equation \eqref{eqn:HFE}.

 \subsection{Decreasing action functional}\label{subsec:dereasing_A}
At this point we provide a lemma showing that the steady state solution of \eqref{eqn:HFE} is a solution of \eqref{eqn:EL}. 
\begin{lemma}[Gradient decent rule under IBCs]\label{lem:decreasing V}
Let $x(t,s)$ be a solution to the AGHF \eqref{eqn:HFE} with an initial condition \eqref{ic} and boundary conditions \eqref{partial bc1}, \eqref{partial bc2}. Then $\frac{\partial \A(x(\cdot,s))}{\partial s}\leq 0$ and it is $0$ if and only if \eqref{eqn:EL} is satisfied on the curve $x(\cdot,s)$.
\end{lemma}
With some modification on the boundary conditions, its proof is essentially the same as the proof of Lemma 1 in \cite{SL19NOLCOS} and provided in the appendix.

As a remark, similar arguments also hold if we require the indefinite boundary states not fully free but in some subsets of $\R^n$. In other words, if the motion planning problem needs to satisfy the boundary condition that
\begin{align*}
    x(0)\in \Omega^i&:=\{x\in\R^n:\phi_j^i(x)=0,j=1,\cdots, n^i\},\\
    x(T)\in \Omega^f&:=\{x\in\R^n:\phi_k^f(x)=0,k=1,\cdots, n^f\},
\end{align*}
then all we need are some transversality conditions such that $\frac{\partial L}{\partial x_t}(x(0,s),x_t(0,s))$ is a linear combination of $\nabla \phi_j^i$'s and $\frac{\partial L}{\partial x_t}(x(T,s),x_t(T,s))$ is a linear combination of $\nabla \phi_k^f$'s in the non-degenerate case. The proof is similar and hence omitted. For more information on transversality conditions for optimization, see, e.g., \cite{CVOC}.

%Nevertheless, This kind of boundary conditions leads to some difficulties in the implementation of the algorithm, as we will later see in the discussion on implementation in \texttt{matlab}.

\subsection{Algorithm}\label{subsec:free_end_states_algorithm}
Our algorithm for motion planning with IBCs and fixed terminal time is very similar to our old algorithm for fixed boundary states, consisting of the following steps:
\begin{enumerate}
\item[S1:] Find a {\it bounded} $n\times (n-m)$ $x$-dependent matrix $F_c(x)$, differentiable in $x$, such that 
\begin{equation}\label{def:bar_F}
\bar F(x):=\begin{pmatrix}F_c(x)|F(x)
\end{pmatrix} \in \R^{n \times n}
\end{equation}
is invertible for all $x\in\R^n$. The matrix $F_c(x)$ can be obtained using, e.g., the Gram-Schmidt procedure.
\item[S2:] Evaluate 
\begin{equation}\label{def:G}
G(x):=(\bar F(x)^{-1})^\top D\bar F(x)^{-1}
\end{equation}
where $D:=\mbox{diag}(\underbrace{\lambda,\cdots,\lambda}_{n-m},\underbrace{1,\cdots,1}_{m})$ for some sufficiently large $\lambda>0$ and set
\begin{equation}\label{def:L}
L(x,\dot x)=(\dot x-h(x))^\top G(x)(\dot x-h(x)).    
\end{equation}

\item[S3:]  
Solve the AGHF~\eqref{eqn:HFE} with boundary conditions~\eqref{partial bc1}, \eqref{partial bc2} and initial condition~\eqref{ic}. Denote the solution by $x(t,s)$;
\item[S4:] For some sufficiently large $\bar s$, Evaluate the \emph{extracted control} by
\begin{equation}\label{control_extraction}
         u(t):=\begin{pmatrix}
        0&I_{m\times m}
        \end{pmatrix}\bar F(x(t,\bar s))^{-1}( x_t(t,\bar s)-h(x(t,\bar s))
   \end{equation}
\end{enumerate}
{\bf Output:} The control $u(\cdot)$ obtained in~\eqref{control_extraction} yields a trajectory $\tilde x(\cdot)$ when integrating~\eqref{sys} from the initial value $\tilde x(0)=x(0,\bar s)$, which is our solution to the motion planning problem. We call it \emph{integrated path}.

Again, as we stated in \cite{SL19NOLCOS}, the integrated path from this algorithm will only be an approximated admissible path and how close the end states to the desired boundary conditions depends on how large are the two parameters $\lambda$ and $\bar s$.  In addition, this algorithm will work only if the motion planning problem is feasible and the initial curve $v$ is chosen in some neighborhood of an admissible path, which is not pre-known. The theoretical result is almost the same as our previous one so we only provide the statements but omit the proof here.  
\begin{theorem}
\label{thm:main}
Consider the system \eqref{sys} and assume the motion planning problem with IBCs \eqref{partial bc1} and fixed terminal time $T$ is feasible. Then there exists $C>0$ such that for any $\lambda>0$, there exists an open set $\Omega_\lambda\subseteq\X'$ so that as long as the initial curve $z\in\Omega_\lambda$, The integrated path $\tilde x(\cdot)$ from our algorithm with sufficiently large $\bar s$ has the properties that for all $(i,0)\in S_{bc}$,
\begin{equation}\label{initialpoint_same}
    \tilde x_i(0)=x^{bc}(i,0)
\end{equation}
and for all $(i,T)\in S_{bc}$,
\begin{equation}\label{endpoint_convergence}
|\tilde x_i(T)-x^{bc}(i,T)|\leq \sqrt{\frac{3TMC}{\lambda}}\exp{\left(\frac{3T}{2}(L_2^2T+L_1^2C)\right)}.
\end{equation}
where $L_1,L_2$ are the Lipschitz constants of $h,F$ and $M$ is the upper bound of $\Vert F_c\Vert$, which is constructed in the algorithm S1. 
\end{theorem}
\subsection{Comments on implementation in \texttt{matlab}}
Our algorithm is implemented in \texttt{matlab}. The first and second step of the algorithm is processed symbolically. Step 3 is implemented via the commend \texttt{pdepe}, where the type of hyperbolic PDEs it is capable of solving is of the form
\begin{equation}\label{eqn:matlab_PDE}
    c(t,s,x,x_t)x_s=t^{-m}\frac{\partial}{\partial t}\left(t^mf(t,s,x,x_t)\right)+s(t,s,x,x_t)
\end{equation}
By comparison with \eqref{eqn:HFE}, we see that in our case we simply need $m=0, c=G, f=\frac{\partial L}{\partial x_t}$ and $s=-\frac{\partial L}{\partial x}$, which can also be done symbolically. In addition, the boundary conditions are formulated in the form
\[
p(t,s,x)+q(t,s)f(t,s,x,x_t)=0
\]
where the $f$ is the same as in \eqref{eqn:matlab_PDE}, and thus is $\frac{\partial L}{\partial x_t}$ in our case. Therefore for the boundary conditions \eqref{partial bc1} and \eqref{partial bc2}, the implementation is simply
\[
\begin{array}{ll}
   p_i(t,s,x)=x_i(t,s)-x^{bc}(i,t),&q_i(t,s)=0,\\
   &\forall (i,t)\in S_{bc},s\geq 0;\\
    p_i(t,s,x)=0,&q_i(t,s)=1,\\
    &\forall (i,t)\in S\backslash S_{bc},s\geq 0.
\end{array}
\]
%However for the more complicated boundary conditions discussed in the end of Section~\ref{subsec:dereasing_A}, there is no obvious way to implement them within this \texttt{pdepe} framework.

\section{Free terminal time}\label{sec:FTT}

For a driftless affine system, If $u(\cdot):[0,1]\to\R^{m}$ gives an admissible path satisfying the boundary condition $x(0)=x^i,x(1)=x^f$, then $\frac{1}{T}u\left(\frac{t}{T}\right)$  gives a time scaled admissible path with $x'(0)=x^i,x'(T)=x^f$. In other words, there is no need to consider the terminal time $T$ other than $1$ since it can always be done by scaling the inputs. However, when there are input constraints or in the presence of the drift term, such scaling is no longer true and hence the minimization of $\A$ in \eqref{def:A} as studied in Lemma~\ref{lem:decreasing V} should be with respect to $T$ as well. In other words, unlike the driftless case where the reachable space of a driftless system is independent of $T$, the reachable space of an affine system with drift or constrained inputs is somehow related to the terminal time $T$. For example, the unicycle with unit linear velocity can only reach planar positions within the ball of radius $T$ at time $T$. Thus we cannot simply just fix a random $T$ prior to minimizing $\A$, in which case solutions may even not exist.  

\subsection{Augmenting the true time}
Compared with the case of fixed terminal time, FTT in the view of maximum principle means that the Hamiltonian is identically $0$ along the optimal trajectory. That information is not helpful in our case, as we do not rely our analysis on the costates nor Hamiltonian. Instead, while we still consider functions defined over a fixed domain $[0,1]$, the way to tackle FTT is to augment a new state $\tau\in \R$ to the system, which is the true time variable that starts from $\tau(0)=0$ and $\tau(1)=T$ yet to be determined. There is also an additional constraint on the function $\tau(\cdot)$ that it needs to be strictly increasing, in which case the inverse function $\tau^{-1}$ exists and we can recover the control as a function of the true time from $u(\cdot)$ by $u^{\dagger}(t)=u(\tau^{-1}(t))$.  For smooth $\tau(\cdot)$, this monotonicity constraint can be resolved by deploying our earlier technique on constrained inputs by treating the derivative of $\tau$ as another extra state, or simply we define $\dot\tau(t)=a(t)^2$, $\dot a(t)=u_0(t)$ where $u_0$ is the additional input to the twice-augmented system. Notice that since $\tau$ is the true time, $\frac{d x}{d\tau}$ should obey the true system dynamics \eqref{sys} instead of $\frac{dx}{dt}$. Thus using chain rule, we have $\dot x:=\frac{dx}{dt}=\frac{dx}{d\tau}\frac{d\tau}{dt}=h(x)a^2+F(x)a^2 u$. In summary, denote the augmented state 
\begin{equation}\label{def:barx}
    x'=\begin{pmatrix}
x\\ \tau \\ a
\end{pmatrix},
\end{equation}
we have
\begin{equation}\label{free_T_augmented_sys}
    \dot {x}'=\begin{pmatrix}
    \dot x\\\dot \tau\\\dot a
    \end{pmatrix}=\underbrace{\begin{pmatrix}
    h(x)a^2\\a^2\\0
    \end{pmatrix}}_{h'(x')}+\underbrace{\begin{pmatrix}
    F(x)a&0\\0&0\\0&1
    \end{pmatrix}}_{F'(x')}\begin{pmatrix}
    au\\u_0
    \end{pmatrix}
\end{equation}
By this augmentation we have new drift term $h'$ and new admissible control direction matrix $F'$. The reason why we take one $a$ out from $F'$ and multiply it to the control $u$ will be explained later when we discuss the total energy consumption of the planned path. In addition, by observation we see that if the inadmissible control direction matrix is constructed by
\begin{equation}\label{def:F'c}
F'_c(x'):=\begin{pmatrix}
F_c(x)&0\\0&1\\0&0
\end{pmatrix}
\end{equation}
then $\bar F'=(F'_c\ F')$ is full rank if $\bar F=(F_c\ F)$ is full rank as we needed earlier. Because the dimension of the system \eqref{free_T_augmented_sys} is now $n+2$, $D',G',L'$ should all be defined accordingly. There are also some small tweaks on the boundary conditions. We still have the old boundary conditions \eqref{partial bc1}, \eqref{partial bc2}; we also have a new boundary condition on $\tau$ because true time also starts at $0$: 
\begin{equation}\label{bc_tau0}
\tau(0,s)=0\quad\forall s\geq 0.
\end{equation}
We do not have any constraints on $\tau(1,s),a(0,s),a(1,s)$; nevertheless, according to the  previous discussion on IBCs we should have
\begin{equation}\label{bc_free}
     \begin{aligned}
     &\frac{\partial L}{\partial \tau_t}(\bar x(1,s),{\bar x}_t(1,s),1)=0,\\
     &\frac{\partial L}{\partial a_t}(\bar x(0,s),{\bar x}_t(0,s),0)=0,\\
     &\frac{\partial L}{\partial a_t}(\bar x(1,s),{\bar x}_t(1,s),1)=0
     \end{aligned}
\end{equation}
for all $s\geq 0$ as a complement in order to solve our AGHF. The rest can be proceeded similarly to the algorithm in Section~\ref{subsec:free_end_states_algorithm} and we summarize it in the following subsection:

\subsection{Algorithm for FTT problem}\label{subsec:new_algorithm}
\begin{enumerate}
    \item[S1:] Augment the states as in \eqref{def:barx}. Find a {\it bounded} $n\times (n-m)$ $x$-dependent matrix $F_c(x)$, differentiable in $x$, such that 
    $\bar F$ defined in \eqref{def:bar_F} is invertible for all $x\in\R^n$. Denote $\bar F'(x')=(F'_c(x)\ F'(x'))$ where $F',F'_c$ come from \eqref{free_T_augmented_sys}, \eqref{def:F'c}.
    % \[
    % \bar F'(x')=\begin{pmatrix}
    % F_c(x)&0&F(x)a&0\\0&1&0&0\\0&0&0&1
    % \end{pmatrix}.
    % \]
    \item[S2:]  Evaluate 
\begin{equation}\label{def:G'}
G'(x'):=(\bar F'(x')^{-1})^\top D'\bar F'(x')^{-1}
\end{equation}
where $D'=\mbox{diag}(\underbrace{\lambda,\cdots,\lambda}_{n-m+1},\underbrace{1,\cdots,1}_{m+1})$ and set
\begin{equation}\label{def:L'}
    L'(x',\dot x')=(\dot x'-h'(x'))^\top G'(x')(\dot x'-h'(x')).
\end{equation}
    \item[S3:] 
    Pick some $T_g>0,a^i_g,a^f_g\in \R$ as the initial guess for $T,a(0),a(1)$. Let $z'\in C^1([0,1]\to\R^{n+2})$ be an initial curve such that $z'_i(t)=x^{bc}(i,t)$ for all $(i,t)\in S_{bc}$, $z'_{n+1}(0)=0, z'_{n+1}(1)=T_g, z'_{n+2}(0)=a^i_g, z'_{n+2}(1)=a^f_g$. Solve the AGHF~\eqref{eqn:HFE} with boundary conditions~\eqref{partial bc1}, \eqref{partial bc2}, \eqref{bc_tau0}, \eqref{bc_free} and initial curve $z'$ described above. Denote the solution by $x'(t,s)$.
    \item[S4:] Fix $\bar s$ sufficiently large. Define 
    \[
    w(t):=\bar F'(x(t,\bar s))^{-1} (x_t'(t,\bar s)-h'(x'(t,\bar s)).
    \]
    Split $w$ so $w^\top =(v^\top\ v_0\  u^\top\ u_0)$ for some $v\in C^0([0,1]\to\R^{n-m}), u\in C^0([0,1]\to \R^m)$, $v_0,u_0\in C^0([0,1]\to \R)$. Define
    \begin{equation}
        \tilde\tau(s)=\int_0^sa(t)^2dt
    \end{equation}
    where 
    \begin{equation}\label{def:a}
        a(t)=\int_0^tu_0(r)dr+x'_{n+2}(0,\bar s)=x'_{n+2}(t,\bar s) ,
    \end{equation}
    then $T=\tilde \tau(1)$ is our resultant terminal time and
    \[
    u^{\dagger}(t)=\frac{u(\tilde \tau^{-1}(t))}{a(\tilde \tau^{-1}(t))}
    \]
    is our extracted control.
\end{enumerate}
{\bf Output:} The integrated path $\tilde x^{\dagger}$ is obtained by integrating \eqref{sys} with the extracted control $u^{\dagger}(\cdot)$ and initial value $\tilde x^{\dagger}(0)=(x'_1(0,\bar s)\ x'_2(0,\bar s)\cdots x'_{n}(0, \bar s))^\top $.

\subsection{Cost minimizing}
We provide a theorem in the FTT case here, which is similar to Theorem~\ref{thm:main}. Its proof is contained in the appendix.
\begin{theorem}
\label{thm:2}
Consider the system \eqref{sys} and assume the motion planning problem with IBCs \eqref{partial bc1} and FTT is solvable. The integrated path $\tilde x^{\dagger}(\cdot)$ from our algorithm with properly chosen initial curve $z'$ and sufficiently large $\bar s$ has the properties that for all $(i,0)\in S_{bc}$,
\begin{equation}\label{initialpoint_same2}
    \tilde x^{\dagger}_{i}(0)=x^{bc}(i,0)
\end{equation}
and there exists $K>0$ such that for all $(i,T)\in S_{bc}$,
\begin{equation}\label{endpoint_convergence2}
|\tilde x^{\dagger}_{i}(T)-x^{bc}(i,T)|\leq \frac{K}{\sqrt{\lambda}}.
\end{equation}
\end{theorem}

Next we provide a heuristic argument that our extracted control is ``economical". Plug $x'(t,\bar s)$ which is derived from Step 3 into $L'$ defined in Step 2, we have
\begin{multline*}
L'(x'(t,\bar s),x'_t(t,\bar s))=w(t)^\top D' w(t)\\
=\lambda|v(t)|^2+\lambda v_0(t)^2+|u(t)|^2+u_0(t)^2\geq |u(t)|^2,  
\end{multline*}
From our analysis of Lemma~\ref{lem:decreasing V} we see that the action functional $\A=\int_0^1L'dt$ is minimized when solving the AGHF; in other words, the $L^2$-norm of $u(\cdot)$ is relatively small from our algorithm. On the other hand, again by the change of variable via $\tilde \tau^{-1}(r)=s$, this $L^2$-norm is exactly the energy of the actual input:
\[
E=\int_0^T|u^{\dagger}(r)|^2dr=\int_0^T\frac{|u(\tilde \tau^{-1}(r))|^2}{a(\tilde \tau^{-1}(r))^2}dr=\int_0^1|u(s)|^2ds
\]
This in fact is not a coincidence; it is only achieved when we design the $F'(x')$ in that particular form as in \eqref{free_T_augmented_sys} where we have shifted one $a$ to the input. As a summary, our algorithm not only finds an approximation to an admissible path which satisfying the IBCs with FTT, the energy consumption of the planned path is also relatively small. 
% Hence for a smooth curve $\bar x$ such that 
% \[
% \dot{\bar x}(t)=h(\bar x(t))l_0(t)+\bar F(\bar x(t))\begin{pmatrix}
% a(t)u(t)\\v(t)
% \end{pmatrix}+\bar G(\bar x(t))l(t)
% \]
% Then with some effort we can show that
% \[
% V(\bar x)=\int_0^1L(\bar x,\dot{\bar x})dt=\int_0^1k((l_0-1)^2+|l|^2)+\frac{1}{k}v^2+|au|^2dt
% \]
% Intuitively, since we have chosen $k$ to be extremely large, minimizing $V$ forces the term $(l_0-1)^2+|l|^2$ to be small; that is, $l_0\approx 1$ and $l\approx 0$. This is the same as what we have practiced in the motion planning problem for system with drift and fixed terminal time. The factor $\frac{1}{k}$ in front of the second term suggests that minimizing $V$ has insignificant effect on the magnitude of $v$; In other words, we don't care much about the value of $v$ in this motion planning problem. As a result, what gets minimized is essentially $\int_0^1|au|^2dt$. But what is this? Recall that when parametrized by the real time $\tau$, the energy cost of the system is given by
% \[
% J_{cost}=\int_0^T|u^{\dagger}(\tau)|^2d\tau
% \]
% Integrating by substitution, we have
% \[
% J_{cost}=\int_0^T|u(\tau^{-1}(t))|^2\frac{d\tau}{dt}dt=\int_0^1|au|^2dt,
% \]
% which is exactly the essential part of $V(\bar x)$. Hence our algorithm with large $k$ gives an path  and an terminal time along which the energy of the system is small.

\section{Case study: unicycle with unit linear velocity}\label{sec:example}

A planar unicycle has three state variables, $x,y$ which represent its planar position and $\theta$ which represents its orientation. The kinematics of a unicycle with unit linear velocity is given by
\begin{equation}\label{uv_unicycle}
    \begin{pmatrix}
    \dot x\\\dot y\\\dot\theta
    \end{pmatrix}=\begin{pmatrix}
    \cos\theta\\\sin\theta\\0
    \end{pmatrix}+\begin{pmatrix}
    0\\0\\1
    \end{pmatrix}u,
\end{equation}
from which we have $h=\begin{pmatrix}\cos\theta&\sin\theta&0\end{pmatrix}^\top$ and $F=\begin{pmatrix}0&0&1\end{pmatrix}^\top$. It is not hard to see that $F_c=\begin{pmatrix}1&0\\0&1\\0&0
\end{pmatrix}$ is the orthogonal complement to $F$. Thus according to our algorithm we have
\[
h'(x')=\begin{pmatrix}
a^2\cos\theta\\a^2\sin\theta\\0\\a^2\\0
\end{pmatrix},\quad\bar F'(x')=\begin{pmatrix}
1&0&0&0&0\\0&1&0&0&0\\0&0&0&a&0\\0&0&1&0&0\\0&0&0&0&1
\end{pmatrix}.
\]

Consider the canonical parallel parking problem that $(x,y,\theta)|_{\tau=0}=(0,0,0)$ and $(x,y,\theta)|_{\tau=T}=(0,1,0)$ where $T$ is to be determined. We would also like to ensure that the energy cost $E=\int_0^Tu(\tau)^2d\tau$ is small. We pick $\lambda=1000,T_g=10,a_{g}^i=a_{g}^f=1$ and Let $z'$ be the line segment connecting the boundary conditions for our algorithm.

\begin{figure}
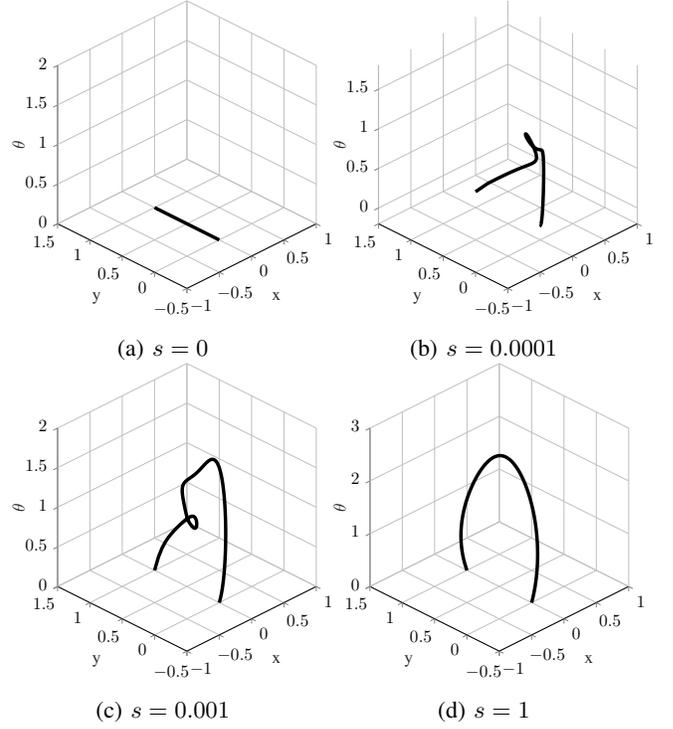

\tikzset{every picture/.style={scale=0.65}}
 \begin{subfigure}{.48\columnwidth}
 \centering
 \input{figures/3D_s=0}
 \caption{$s=0$}\label{subfig:s=0}
 \end{subfigure}
  \begin{subfigure}{.48\columnwidth}
 \centering
 \input{figures/3D_s=e-4}
 \caption{$s=0.0001$}\label{subfig:s=e-4}
 \end{subfigure}
 
  \begin{subfigure}{.48\columnwidth}
 \centering
 \input{figures/3D_s=e-3}
 \caption{$s=0.001$}\label{subfig:s=e-3}
 \end{subfigure}
  \begin{subfigure}{.48\columnwidth}
 \centering
 \input{figures/3D_s=1}
 \caption{$s=1$}\label{subfig:s=1}
 \end{subfigure}
 \caption{The $(x,y,\theta)$ plots of the solution of \eqref{eqn:HFE} for different values of $s$.}\label{fig:curve_evolution}
\end{figure}

\begin{figure}
\tikzset{every picture/.style={scale=0.8}}
 \centering
 \input{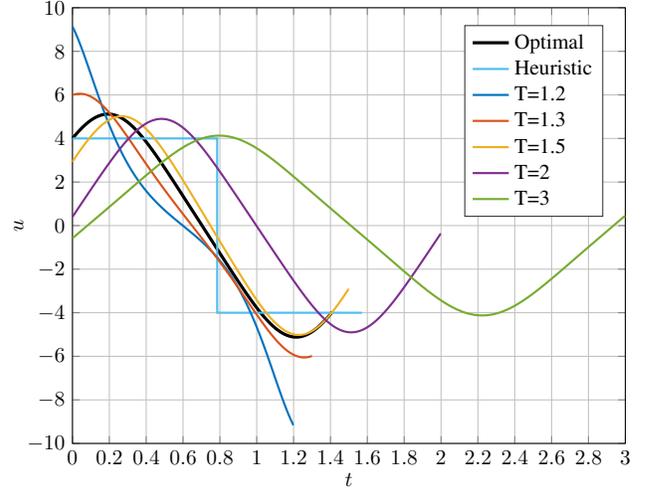}
 \caption{The extracted control for the FTT algorithm, the heuristic control and the extracted controls from our previous algorithm with fixed $T$.}\label{fig:resultant_control}
 \end{figure}

\begin{figure}
 \tikzset{every picture/.style={scale=0.8}}
 \centering
 \input{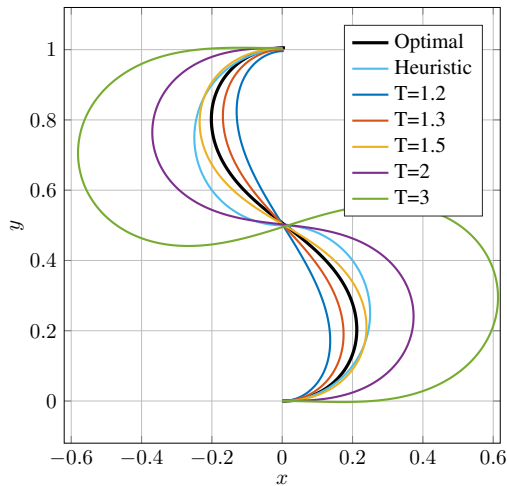}
 \caption{The integrated path for the FTT algorithm, the heuristic path and the integrated paths from our previous algorithm with fixed $T$.}\label{fig:resultant_path}
\end{figure}

The evolution of the $(x,y,\theta)$ coordinates in the AGHF solution $x'(s,t)$ with respect to $s$ is shown in Figure~\ref{fig:curve_evolution}. Notice that in Figure~\ref{subfig:s=0} we have $s=0$ and hence it is indeed the first three coordinates of the initial curve $z'$. In Figure~\ref{subfig:s=1}, the curve barely changes any more with respect to $s$ so the PDE solver is stopped and we use $\bar s=1$ to extract the control. The extracted control is shown as the black curve in Figure~\ref{fig:resultant_control} and integrated path is shown as the black curve in Figure~\ref{fig:resultant_path}. It turns out that by our algorithm $T=1.4072$ and $E=21.1022$. As a comparison, a heuristic admissible path for the unicycle system \eqref{uv_unicycle} which consists of two semicircles is considered. Such a path has a total length of $\frac{\pi}{2}$, and hence the total traveling time is also $\frac{\pi}{2}$ because of unit velocity. By observation we see that $u$, or the turning rate, is equal to the curvature of the path and thus $u(t)=4$ for the first half and $u(t)=-4$ for the second half. As a result, the total energy cost in this case is $4^2\times\frac{\pi}{2}\approx 25.13$. Both the total time and energy cost of this heuristic path is larger than what we derived from our proposed algorithm. In addition, we also applied our motion planning algorithm studied in ~\cite{SL19NOLCOS} for fixed terminal time with varies $T$. Their extracted controls and integrated paths are also illustrated in Figure~\ref{fig:resultant_control} and Figure~\ref{fig:resultant_path}.

\begin{figure}
 \tikzset{every picture/.style={scale=0.75}}
 \centering
 \input{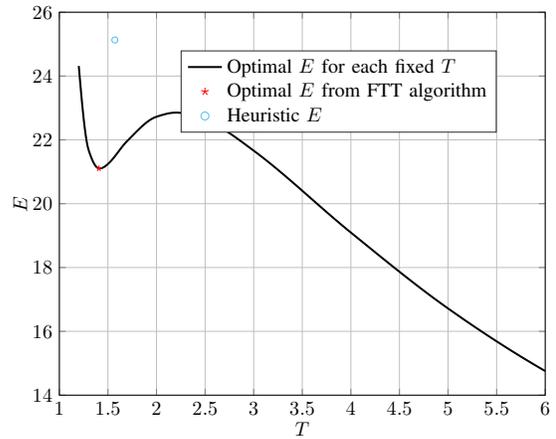}
 \caption{Energy vs. $T$.}\label{fig:energy_comparison}
\end{figure}

The energy costs of these different results are shown in Figure~\ref{fig:energy_comparison}. Notice that this motion planning problem has no solution with global minimal energy cost. This is because the parallel parking of the constant linear velocity unicycle can always be accomplished by an S-shaped curve with arbitrarily large turning radius $r$, as seen by the trend of solutions with larger $T$ in Figure~\ref{fig:resultant_path}. By doing that the magnitude to the control is approximately $O(1/r)$ while the total time is approximately $O(r)$ and hence $E=\int_0^T|u|^2dt=O(1/r)\to 0$ as $r$ increases to infinity. This asymptotic behavior is also reflected on the plot of energy in Figure~\ref{fig:energy_comparison}. However finding globally optimal or sub-optimal solution for the motion planning is out of our interest because it requires infinite or extremely long travel time of the unicycle and hence impractical. On the other hand, our FTT algorithm is able to produce a value of $T$ which gives us a local optimal solution; in addition, this $T$ and energy cost coincide with the local optimal result generated by iterations of the motion planning algorithm with fixed terminal time.

As a final remark, we want to comment on finding the true optimal solution to this motion planning problem via maximum principle. We start by formulating the Hamiltonian as
\begin{equation}\label{def:Hamiltonian}
    H=p^\top f-L=p_1\cos(\theta)+p_2\sin(\theta)+p_3u-u^2
\end{equation}
Now by maximum principle we have $u=\frac{p_3}{2}$, $p_1,p_2$ are constants and $\dot p_3=p_1\sin(\theta)-p_2\cos(\theta)$.
% \[
% \begin{array}{ccc}
% \frac{\partial H}{\partial u}=0      &\Rightarrow& u=\frac{p_3}{2},\\
% \dot p=-\frac{\partial H}{\partial x} &\Rightarrow& p_1,p_2 \mbox{ are constants, }\dot p_3=p_1\sin(\theta)-p_2\cos(\theta).
% \end{array}
% \]
In addition, since the problem has a free terminal time, we have $H\equiv 0$. While we still have boundary conditions on the states $x,y,\theta$, We do not know the value of $p_1,p_2$ and there are neither any boundary conditions for $p_3$. On the other hand, if we take the second derivative of $\theta$, we see that
\begin{equation}\label{ddtheta}
    \ddot \theta=\frac{du}{d\theta}=\frac{1}{2}\frac{dp_3}{dt}=\frac{p_1}{2}\sin(\theta)-\frac{p_2}{2}\cos(\theta),
\end{equation}
which is a second order nonlinear ODE. While there are no general solutions for nonlinear ODEs with IBCs and FTT, the general solution of \eqref{ddtheta} can be expressed in terms of Jacobi elliptic functions \cite{whittaker_watson_1996}, which requires quite a lot of work. It can also be seen that finding the exact optimal path via maximum principle is difficult to be generalized for more complicated systems or in higher dimensions. On the contrary, in trade of the true minimal cost and the accuracy of the exact expression of the optimal solution, our algorithm is quite systematic and usually gives a good approximating solution within reasonable amount of computation time.

\section{Conclusion}\label{sec:conclusion}
 In this paper we have extended our earlier method for motion planning for affine control systems with fixed boundary conditions and fixed terminal time to indefinite boundary conditions and free terminal time. We have first shown that the deficiency of boundary conditions can be completed by constraints on the Lagrangian with respect to the derivative of the corresponding states and it was verified via a perturbation argument. On the other hand, a time scaling has been applied to the motion planning problem, which resulted in an augmented affine system with drift and indefinite boundary conditions and hence the motion with indefinite terminal time can be planned by appealing to the techniques we developed in the previous work and the analysis on indefinite boundary conditions. In the end of the paper we have also studied a canonical example of unicycle with constant linear velocity and used our algorithm to show how parallel parking can be accomplished in the most economical way.

\appendix
\subsection{Proof of Lemma~\ref{lem:decreasing V}}
Firstly, by first order approximation we have
\[
x(t,s+\delta)=x(t,s)+\delta x_s(t,s)+o(\delta)
\]
Plug it into the first order variation of $L$, we have
\begin{align*}
    &\A(x(\cdot,s+\delta))=\int_0^TL(x(t,s+\delta),x_t(t,s+\delta))dt\\
    &=\int_0^TL(x(t,s),x_t(t,s))+(\delta x_s(t,s)+o(\delta))^\top \frac{\partial L}{\partial x}\\
    &\quad+\left(\frac{d}{dt}(\delta x_s(t,s)+o(\delta))\right)^\top \frac{\partial L}{\partial x_t}+o(\delta)dt\\
    &=\A(x(\cdot,s))+\delta\int_0^Tx_s(t,s)^\top\frac{\partial L}{\partial x}+x_{ts}(t,s)^\top\frac{\partial L}{\partial x_t}dt+o(\delta),
\end{align*}
where all $o(\delta)$ terms are collected together. Use integration by parts for the $x_{ts}(t,s)^\top\frac{\partial L}{\partial x_t}$ term, we have
\begin{multline*}
\A(x(\cdot,s+\delta))=\A(x(\cdot,s))+\delta\left(\left.x_s(t,s)^\top\frac{\partial L}{\partial x_t}\right|_0^T\right.\\
\left.+\int_0^Tx_s(t,s)^\top\frac{\partial L}{\partial x}-x_s(t,s)^\top\frac{d}{dt}\frac{\partial L}{\partial x_t}dt\right)+o(\delta).    
\end{multline*}
Our new boundary conditions \eqref{partial bc1}, \eqref{partial bc2} imply that $x_s^\top\frac{\partial L}{\partial x_t}=0$ for both $t=0,T$ and all $s\geq 0$. Hence the integrated term $\left.x_s(t,s)^\top\frac{\partial L}{\partial x_t}\right|_0^T$ vanishes. Thus plug in the AGHF \eqref{eqn:HFE} here, 
\begin{multline*}
\Delta \A=\delta \int_0^Tx_s(t,s)^\top\left(\frac{\partial L}{\partial x}-\frac{d}{dt}\frac{\partial L}{\partial x_t}\right)dt+o(\delta)\\
=-\delta\int_0^TG(x)|x_s(t,s)|^2dt+o(\delta)  
\end{multline*}
where $\Delta \A=\A(x(\cdot,s+\delta))-\A(x(\cdot,s))$ and hence
\[
    \frac{\partial \A(x(\cdot,s))}{\partial s}=\lim_{\delta\to 0}\frac{\Delta A}{\delta}=-\int_0^TG(x)|x_s(t,s)|^2dt\leq 0
\]
and equality is achieved if and only if $x_s(t,s)=0$ almost everywhere for $t\in[0,T]$. Because of \eqref{eqn:HFE} and the fact that $x(\cdot,s)\in C^1$, \eqref{eqn:EL} is satisfied on $x(\cdot,s)$.  
\hfill$\square$

\subsection{Proof of Theorem~\ref{thm:2}}
Note the property \eqref{initialpoint_same2} is directly given by the construction. The major difference between the two algorithms is in Step 4, where a time scaling is involved in the second algorithm on FTT. Nevertheless, if we directly feed $(u,u_0)$ to the system \eqref{free_T_augmented_sys} as required by the Step 4 in the first algorithm, by the results of Theorem~\ref{thm:main} we again conclude the bounds of \eqref{endpoint_convergence} for all $(i,T)\in S_{bc}$, where
\[
\tilde x(t)=\tilde x^{\dagger}(0)+\int_0^th(\tilde x(s))a(s)^2+F(\tilde x(s))a(s)u(s)ds
\]
and $a$ as defined in \eqref{def:a}. On the other hand, the integrated path is given by
\begin{multline*}
\tilde x^{\dagger}(t)=\tilde x^{\dagger}(0)+\int_0^t h(\tilde x^{\dagger}(r))+F(\tilde x^{\dagger}(s))u^{\dagger}(r)dr\\
=    \tilde x^{\dagger}(0)+\int_0^t h(\tilde x^{\dagger}(r))+\frac{F(\tilde x^{\dagger}(r))}{a(\tilde \tau^{-1}(r))}u(\tilde \tau^{-1}(r))dr
\end{multline*}
Set $\tilde \tau^{-1}(r)=s$ and notice that $\frac{dr}{ds}=\frac{d\tilde \tau(s)}{ds}=a(s)^2$, we see that
\begin{small}\begin{multline*}
\tilde x^{\dagger}(t)=\tilde x^{\dagger}(0)+\int_0^{\tilde \tau^{-1}(t)} \bigg(h(\tilde x^{\dagger}(\tilde \tau(s)))\\+\frac{F(\tilde x^{\dagger}(\tilde \tau(s)))}{a(s)}u(s)\bigg)a(s)^2ds\\
=    \tilde x^{\dagger}(0)+\int_0^{\tilde \tau^{-1}(t)} h(\tilde x^{\dagger}(\tilde \tau(s)))a(s)^2+F(\tilde x^{\dagger}(\tilde \tau(s)))a(s)u(s)ds
\end{multline*}
\end{small}
which implies that $\tilde x^{\dagger}(t)=\tilde x(\tilde \tau^{-1}(t))$. In particular, $\tilde x^{\dagger}(T)=\tilde x(1)$ and we conclude \eqref{endpoint_convergence2} from \eqref{endpoint_convergence}.
\hfill$\square$

\bibliographystyle{IEEEtran}
\bibliography{reference}

\end{document}